\documentclass[journal]{IEEEtran}
\ifCLASSINFOpdf
\else
\fi

\usepackage{dsfont}
\usepackage{amsfonts}
\usepackage{mathrsfs}
\usepackage{}
\usepackage{colortbl}
\usepackage{amssymb,amsmath,graphicx}
\usepackage{caption}

\usepackage{subfigure}
\usepackage{float}

\newtheorem{Remark}{\bf Remark}[section]

\newtheorem{Problem}{\bf Problem}[section]

\newenvironment{Proof}{\noindent{\em Proof:\/}}{\hfill $\Box$\par}

\newtheorem{Theorem}{\bf Theorem}[section]

\newtheorem{Lemma}{\bf Lemma}[section]

\newtheorem{Assumption}{\bf Assumption}[section]

\newcommand{\EQ}{\begin{eqnarray}}

\newcommand{\EN}{\end{eqnarray}}

\newcommand{\EQQ}{\begin{eqnarray*}}

\newcommand{\ENN}{\end{eqnarray*}}

\begin{document}
\title{An updated version of ``Leader-following consensus for linear multi-agent systems via asynchronous sampled-data control," IEEE Transactions on Automatic Control, DOI: 10.1109/TAC.2019.2948256.}

\author{Wei~Liu,~\IEEEmembership{Member,~IEEE}~and~Jie~Huang,~\IEEEmembership{Fellow,~IEEE}% <-this % stops a space
%\thanks{This work has been supported in part by the Research Grants Council of the
%Hong Kong Special Administration Region under grant No. 412609, and
%in part by the National Natural Science Foundation of China under
%grant No. 61174049.}% <-this % stops a space
\thanks{This work was supported in part by the Research Grants Council of the Hong Kong Special Administration Region under Grant No. 14219516, and in part by National Natural Science Foundation of
China under Projects 61633007 and 61973260. Corresponding author: Jie Huang.} %Tel. +852-39438473. Fax +852-26036002.}
\thanks{Wei Liu and Jie Huang  are with the Department of Mechanical and Automation
Engineering, The Chinese University of Hong Kong, Shatin, N.T., Hong
Kong. E-mail: wliu@mae.cuhk.edu.hk, jhuang@mae.cuhk.edu.hk}
}

\maketitle


%%%%%%%%%%%%%%%%%%%%%%%%%%%%%%%%%%%%%%%%%%%%%%%%%%%%%%%%%%%%%%%%%%%%%%%%%%%%%%%%
\begin{abstract}
In this article, we update the reference \cite{LiuHuang2019a}  in two aspects. First, we note that in order for the control law (12) in \cite{LiuHuang2019a} to be equivalent to the control law (3) in \cite{LiuHuang2019a},  we need to assume that the samplings for all subsystems must be synchronous, i.e.,  we need to assume that $T_{i}=T$ for all $i=1,\cdots,N$. Second, we extend our results from periodic sampling to  aperiodic  sampling.
\end{abstract}

% \fi
\IEEEpeerreviewmaketitle

% or


%%%%%%%%%%%%%%%%%%%%%%%%%%%%%%%%%%%%%%%%%%%%%%%%%%%%%%%%%%%%%%%%%%%%%%%%%%%%%%%%
\section{Introduction}
\IEEEPARstart{O}{ver} the  years, cooperative control of multi-agent systems has attracted extensive attention from the control community. This is due to its wide range of applications in various engineering areas such as coordination of mobile robots, formation of unmanned vehicles, and synchronization of multiple spacecraft systems. The objective of cooperative control is to design a control law using only the information of the neighboring agents to achieve a collective behavior in the overall multi-agent system. Such a control law is called a distributed law.
A fundamental cooperative control problem is called consensus. Depending on whether there is a leader system, the problem can be classified into two types: leaderless consensus and leader-following consensus. The leaderless consensus problem aims to  make the states of a group of agents converge to a same trajectory, while the leader-following consensus problem further requires
the states of a group of follower systems asymptotically track a prescribed trajectory produced by a so-called leader system. So far, both  consensus problems have been widely studied. For example,
 the leaderless consensus problem has been studied in \cite{Munz2011,Olfati2004,RenW2005,Tuna2008},  the leader-following consensus problem has been studied in \cite{HongY2008,HuJ2007,NiW2010},  and  both problems have been studied in \cite{Jadbabaie2003,SuHuang2012}.

%Note that the global behavior of a multi-agent system is jointly determined by two significant factors: the agent dynamics and the communication network topology. First, according to different agent dynamics, the consensus problem has been widely studied for single integrator systems \cite{Jadbabaie2003,Munz2011,Olfati2004}, double integrator systems \cite{HongY2008,HuJ2007,RenW2008} and general linear systems \cite{NiW2010,SuHuang2012,Tuna2008}.
%Second, according to different communication network topologies, the the consensus problem has been widely studied for static network \cite{RenW2008,Tuna2008}, every-time connected switching network \cite{HongY2008,Olfati2004}, %\cite{Hong1,Hu1,Olfati1},
%frequently connected switching network \cite{Vengertsev1,WangJ2008}, and jointly connected switching network \cite{Jadbabaie1,SuHuang2012}. %\cite{Jadbabaie1,Munz1,Ni1,SH3}.

It is noted that most existing results on continuous-time multi-agent systems assume that the information is transmitted continuously and the control laws are also in the continuous-time form.   However, many advanced communication networks only permit digital information transmission, and more and more practical controllers are implemented in digital platforms. Hence, it is more practical to take into account both digital information transmission and digital control laws. The sampled-data control approach has been a most commonly used method for implementing a continuous-time control law in a digital platform \cite{Astrom1997,ChenT1995,Franklin1998}, and recently, this approach
has also been used to address the consensus problem. For example,  in \cite{XieG2009a,XieG2009b}, the sampled-data leaderless consensus problem (SDLLCP) for single-integrator multi-agent systems was studied for the static network case and the switching network case, respectively.
 The SDLLCP was further studied for single-integrator multi-agent systems in \cite{XiaoF2008} and double-integrator multi-agent systems in \cite{GaoY2011}, where the communication networks are assumed to be switched and jointly connected.
 %and the control laws are asynchronous in the sense that the sampling period of each agent is independent of the others'.
 %Also, the time delay issue was considered in \cite{XiaoF2008}.
 In \cite{YuW2011}, a control protocol depending on the sampled position data was proposed to solve the SDLLCP for double-integrator multi-agent systems. Reference \cite{GuanZ2012} further studied the SDLLCP for double-integrator multi-agent systems based on the impulsive control strategy. In \cite{ZhangW2017,ZhangY2010}, the sampled-data leaderless mean square consensus problem was studied for the general linear multi-agent systems with packet losses. Reference \cite{ZhangX2017} studied the SDLLCP for general linear multi-agent systems with switching topologies using the input delay method.
In \cite{TangZ2011}, the sampled-data leader-following consensus problem (SDLFCP) was studied for a class of multi-agent systems by using the direct discretization method, where the follower systems had the single-integrator dynamics and the leader system had the double-integrator dynamics. In \cite{Park2017}, two weighted consensus tracking protocols via computing the network centrality were proposed to solve the SDLFCP for double-integrator multi-agent systems. Reference \cite{XieD2015} studied the bounded SDLFCP for double-integrator multi-agent systems, and the tracking errors were guaranteed to be ultimately bounded.
 In \cite{DingL2013}, a delay-dependent stability criterion was derived to solve the SDLFCP for general linear multi-agent systems. However,
 %Reference \cite{Rakkiyappan2015} studied the sampled-data leader-following mean square consensus problem for general linear multi-agent systems with randomly missing data. In \cite{ZhangD2018}, the sampled-data leader-following mean square consensus problem for general linear multi-agent systems with DoS attack was further studied.
the solvability conditions of the problem in  \cite{DingL2013} depend on the solvability conditions of some linear matrix inequalities.
More results on the sampled-data consensus problem can be found in the recent survey paper \cite{GeX2018} and the references therein.

In this paper, we further study the SDLFCP for general linear multi-agent systems. Compared with the existing results,
we derive solvability conditions of the problem based on rigorous Lyapunov analysis. Specifically,  we give an explicit upper bound for the sampling intervals that guarantees the stability and performance of the closed-loop system  as long as all the sampling intervals are smaller than this upper bound. In addition, our results have some other new features.
%First,  the sampled-data control law is updated asynchronously, which is more practical than the synchronous sampled-data control law since it may be difficult to guarantee synchrony in practice due to the environment disturbances.
 First, we treat  general linear multi-agent systems, which contain single-integrator multi-agent systems and double-integrator multi-agent systems as special cases. Second, our approach applies to both static directed networks and  switching directed networks. Third, we consider aperiodic sampling, which contains periodic sampling as a special case.

\textbf{Notation:}
Denote $\mbox{col}(x_1,...,x_s)=[x_1^T,...,x_s^T]^T$, where $x_i$, $i=1,...,s$, are some column vectors. $\mathbb{Z}^{+}$ denotes the set of all positive integers.
$\mathbb{N}=\{0,\mathbb{Z}^{+}\}$.  $\mathbb{R}^{+}$ denotes the set of all positive real numbers.
 $\|\cdot\|$ denotes the Euclidean norm of a vector or the induced Euclidean norm of a matrix.
Denote the base of the natural logarithm by $\mathbf{e}$.
$\lambda_{\max}(A)$ and $\lambda_{\min}(A)$ denote the maximum eigenvalue and the minimum eigenvalue of a symmetric real matrix $A$, respectively. A matrix $M\in\mathbb{R}^{N\times N}$ is called an $\mathcal{M}$-matrix, if all of its non-diagonal
elements are non-positive and all of its eigenvalues have positive real parts.  For simplicity, we use $x$ to denote $x(t)$ when no ambiguity occurs in this paper.

\section{Preliminaries and Problem formulation}\label{PF}
Consider a class of general linear multi-agent systems composed of $N$ follower systems and a leader system. The dynamics of each follower system is described as follows:
\begin{equation}\label{Follower1}
\begin{split}
%  \dot{z}_{i} =& f_{0i}(z_{i},x_{1i},v,w),\\
\dot{x}_{i}=Ax_{i}+Bu_{i}, ~i=1,\cdots,N
\end{split}
\end{equation}
where $x_{i}\in\mathbb{R}^{n}$ and $u_{i}\in \mathbb{R}^{m}$ are the state and the input of agent $i$, $A\in\mathbb{R}^{n\times n}$ and $B\in \mathbb{R}^{n\times m}$ are two constant matrices.  The dynamics of the leader system is described as follows:
\begin{equation}\label{Leader1}
\begin{split}
\dot{x}_{0}=Ax_{0}
\end{split}
\end{equation}
where $x_{0}\in\mathbb{R}^{n}$ is the state of the leader system.

% Let $N_{\sigma}(t,\tau)$ denote the number of discontinuities of
%$\sigma$ in the open interval $(\tau,t)$ for $t\geq\tau\geq0$. If $\sigma$ satisfies
%\begin{equation}\label{dwelltime1}
%N_{\sigma}(t,\tau)\leq N_{0}+\frac{t-\tau}{\tau_{d}}
%\end{equation}
%for  some positive constants $\tau_{d}$ and $N_{0}$,
%then $\sigma$ is said to possess the property of average dwell-time
%$\tau_{d}$ with chatter bound $N_{0}$. Denote  the set of all signals having this
%property by $S_{ave}[\tau_{d},N_{0}]$. % Note that $S[\tau_{d}]\subset S_{ave}[\tau_{d},N_{0}]$.

Given the multi-agent system composed of \eqref{Follower1} and \eqref{Leader1} and a piecewise constant switching signal $\sigma:[0,\infty)\rightarrow\mathcal {P}$ with $\mathcal {P} = \{1,2,
\cdots, n_{0} \}$, we can define a time-varying digraph  $\bar{\mathcal{G}}_{\sigma(t)}=(\bar{\mathcal{V}},\bar{\mathcal{E}}_{\sigma(t)})$, where   $\bar{\mathcal{V}}=\{0,1,\dots,N\}$ denotes the node set and
$\bar{\mathcal{E}}_{\sigma(t)}\subseteq \bar{\mathcal{V}}\times
\bar{\mathcal{V}}$ denotes the edge set.
%where the node $0$ is associated with the leader system  and the node $i$, $i = 1,\dots,N$,
%is associated with the $i$th follower system.
For $i=1,\dots,N$, $j=0,1,\dots,N$, and $i\neq j$, $(j,i) \in
\bar{\mathcal{E}}_{\sigma(t)}$ if and only if  $u_i$ can use the
information of agent $j$ for control at time $t$. The edge $(i,j)$ is called undirected if $(i,j)\in\bar{\mathcal{E}}_{\sigma(t)}$ implies $(j,i)\in\bar{\mathcal{E}}_{\sigma(t)}$. The digraph $\bar{\mathcal{G}}_{\sigma(t)}$ is called undirected if all edges in $\bar{\mathcal{E}}_{\sigma(t)}$ is undirected.  If the digraph  $\bar{\mathcal{G}}_{\sigma(t)}$ contains a sequence of the edges $(i_{1},i_{2}),(i_{2},i_{3}),\cdots,(i_{k-1},i_{k})$, then node $i_{k}$ is said to be reachable from node $i_{1}$.
Let $\bar{\mathcal{A}}_{\sigma(t)} =[\bar{a}_{ij} (t)]\in \mathbb{R}^{(N+1)\times
(N+1)}$ denote the adjacency matrix of $\bar{\mathcal{G}}_{\sigma(t)}$,  where $\bar{a}_{ii} (t)=0$ and $\bar{a}_{ij} (t)=1\Leftrightarrow (j,i)\in\mathcal{\bar{E}}_{\sigma (t)}$ for $i,j=0,1,\cdots,N$.
Let $\bar{\mathcal{N}}_i(t)=\{j,(j,i)\in \bar{\mathcal{E}}_{\sigma(t)}\}$
denote the neighbor set of agent $i$ at time $t$. Let $H_{\sigma(t)}=[h_{ij}(t)]\in\mathbb{R}^{N\times N}$ with $h_{ij}(t)=-\bar{a}_{ij}(t)$ for $i,j=1,\cdots,N$ and $i\neq j$, and $h_{ii}(t)=\sum_{j=0}^{N}\bar{a}_{ij}(t)$ for $i=1,\cdots,N$.
A digraph $\mathcal{G}_{\sigma(t)}=(\mathcal{V},\mathcal{E}_{\sigma(t)})$ is called a subgraph of $\bar{\mathcal{G}}_{\sigma(t)}=(\bar{\mathcal{V}},\bar{\mathcal{E}}_{\sigma(t)})$ if $\mathcal{V}\subseteq\bar{\mathcal{V}}$ and $\mathcal{E}_{\sigma(t)}\subseteq\bar{\mathcal{E}}_{\sigma(t)}\cap(\mathcal{V}\times\mathcal{V})$ for all $t\geq0$. Note that when $\sigma(t)$ is a constant signal, the communication network becomes a static network. For the static network case, we use $\bar{\mathcal{G}}$,  $\mathcal{G}$, $\bar{\mathcal{A}}$ and $H$ to denote $\bar{\mathcal{G}}_{\sigma(t)}$,  $\mathcal{G}_{\sigma(t)}$, $\bar{\mathcal{A}}_{\sigma(t)}$  and $H_{\sigma(t)}$ for simplicity.
The digraph $\bar{\mathcal{G}}_{\sigma(t)}$
is static if $\bar{\mathcal{E}}_{\sigma(t)}  =  \bar{\mathcal{E}}_{\sigma(0)}
$ for all $t \geq 0$.

Next, we consider the following control law
%\begin{equation}\label{u1}
%\begin{split}
%u_{i}(t)\!=\!K\!\sum_{j=0}^{N}\bar{a}_{ij}(t_{s})(x_{j}(t_{s})\!-\!x_{i}(t_{s})),~\forall t\in[t_{s},t_{s+1})
%\end{split}
%\end{equation}
%where $i=1,\cdots,N$, $t_{s}=sT$, $s\in\mathbb{N}$, $T$ denotes the sampling period, $K$ is a constant matrix with proper dimension.
%%As in \cite{GaoY2011}, it is assumed that the sampling period of each agent is independent of the others' and $T_{i}=l_{i}h$ for $i=1,\cdots,N$, where $l_{i}\in\mathbb{Z}^{+}$ and $h\in\mathbb{R}^{+}$.
\begin{equation}\label{u1}
\begin{split}
u_{i}(t)\!=\!K\!\sum_{j=0}^{N}\bar{a}_{ij}(t_{s})(x_{j}(t_{s})\!-\!x_{i}(t_{s})),~\forall t\in[t_{s},t_{s+1})
\end{split}
\end{equation}
where $i=1,\cdots,N$, $t_{0}=0$, $t_{s+1}=t_{s}+T_{s}$, $s\in\mathbb{N}$,  $T_{s}\in[\underline{T},\bar{T}]$ with $\underline{T}\leq\bar{T}$ being two positive real numbers, and $K$ is a constant matrix with proper dimension.

\begin{Remark}\label{RemarkControlLaw1}
The control law \eqref{u1} is called a distributed sampled-data state feedback control law, since agent $i$ can only make use of the sampled states of its neighbors and itself for feedback control.
%If we let $T_{s}=T$ for all $s\in\mathbb{N}$, then the control law \eqref{u1} is called a  distributed periodic sampled-data state feedback control law, which is a special case of the distributed aperiodic sampled-data state feedback control law.
In fact, the control law \eqref{u1} is motivated by sampling the continuous-time control laws used in \cite{NiW2010,SuHuang2012}. Other similar sampled-data control laws can also be found in the recent survey paper \cite{GeX2018}.
\end{Remark}

We describe the sampled-data leader-following consensus problem as follows:

\begin{Problem}\label{Problem1}
 Given the multi-agent system composed of \eqref{Follower1} and \eqref{Leader1}, and a switching digraph $\bar{\mathcal{G}}_{\sigma(t)}$, design a control law of the form \eqref{u1} with appropriate sampling intervals $T_{s}$, $s\in\mathbb{N}$, such that, for any initial conditions $x_{i}(0)$, $\lim_{t \to \infty}(x_{i}(t)-x_{0}(t))=0$ for $i=1,\cdots,N$.
\end{Problem}

To solve Problem \ref{Problem1}, we introduce the following assumption.
\begin{Assumption}\label{Ass1}
The pair $(A,B)$ is stabilizable.
\end{Assumption}

\begin{Remark}\label{RemarkAss1}
Assumption \ref{Ass1} is a standard assumption for the consensus problem of general linear multi-agent systems, which has also been used in \cite{DingL2013,NiW2010,Tuna2008,ZhangX2017} etc.
\end{Remark}

\section{A Technical Lemma}\label{Lemma}
In this section, we will establish a technical lemma as follows.

\begin{Lemma}\label{Lemma1}
Suppose $W(t):[0,\infty)\rightarrow[0,\infty)$ is  continuous, and there exists a sequence $\{t_{s}: s\in\mathbb{N}, t_{s} \in [0,\infty)\}$ satisfying $t_{s+1}-t_{s}\geq h$ for all $s\in\mathbb{N}$ and some positive real number $h$ such that $W(t)$ is differentiable on each interval $[t_{s},t_{s+1})$  and
 \begin{equation}\label{dotWt1}
\begin{split}
\dot{W}(t)\leq&-\beta_{1}W(t)+\beta_{2}W(t_{s}),~\forall t\in[t_{s},t_{s+1})\\
\end{split}
\end{equation}
where $\beta_{1}$ and $\beta_{2}$ are two positive real numbers with $\beta_{2}<\beta_{1}$. Then
 \begin{equation}\label{Wt1}
\begin{split}
\lim_{t\rightarrow\infty}W(t)=0.\\
\end{split}
\end{equation}
\end{Lemma}
\begin{Proof}
First, if $W(t_{s})=0$ for some $s\in\mathbb{N}$, then, by \eqref{dotWt1} and the fact that $W(t)\geq0$ for all $t\geq0$, we have $W(t)=0$ for all $t\geq t_{s}$. Thus \eqref{Wt1} holds.

Second, consider the case where $W(t_{s})\neq0$ for all $s\in\mathbb{N}$.
%Note that
% \begin{equation}\label{Wts1}
%\begin{split}
%\beta_{2}\sqrt{W(t)W(t_{s})}\leq\frac{\beta_{1}}{2}W(t)+\frac{\beta_{2}^{2}}{2\beta_{1}}W(t_{s}).
%\end{split}
%\end{equation}
%Then, from \eqref{dotWt1} and \eqref{Wts1}, for any $t\in[t_{s},t_{s+1})$,  we have
% \begin{equation}\label{dotWt2}
%\begin{split}
%% \nonumber to remove numbering (before each equation)
%\dot{W}(t)\leq&-\beta_{1}W(t)+\frac{\beta_{1}}{2}W(t)+\frac{\beta_{2}^{2}}{2\beta_{1}}W(t_{s})\\
%=&-\frac{\beta_{1}}{2}W(t)+\frac{\beta_{2}^{2}}{2\beta_{1}}W(t_{s}).\\
%\end{split}
%\end{equation}
For any $ t \in [t_s, t_{s+1})$, solving \eqref{dotWt1} gives
\begin{equation}
\begin{split}
W(t) \leq &\textbf{e}^{-\beta_{1}(t-t_{s})}W(t_{s})+\int_{t_{s}}^{t}\textbf{e}^{-\beta_{1}(t-\tau)}\beta_{2}W(t_{s})d\tau\\
=&\big(\textbf{e}^{-\beta_{1}(t-t_{s})}+\beta_{2}\textbf{e}^{-\beta_{1}t}\int_{t_{s}}^{t}\textbf{e}^{\beta_{1}\tau}d\tau\big)W(t_{s}) \\
%=& \textbf{e}^{-\beta_{1}(t-t_{s})}+\frac{\beta_{2}}{\beta_{1}}\textbf{e}^{-\beta_{1}t}(\textbf{e}^{\beta_{1}t}-\textbf{e}^{\beta_{1}t_{s}})\\
=&\big(\textbf{e}^{-\beta_{1}(t-t_{s})}+\frac{\beta_{2}}{\beta_{1}}(1-\textbf{e}^{-\beta_{1}(t-t_{s})})\big)W(t_{s})\\
=&\big((1-\frac{\beta_{2}}{\beta_{1}})\textbf{e}^{-\beta_{1}(t -t_{s})}+\frac{\beta_{2}}{\beta_{1}}\big)W(t_{s}).\\
%=&\textbf{e}^{-\beta_{1}\bar{T}}+\frac{\beta_{2}}{\beta_{1}}(1-\textbf{e}^{-\beta_{1}\bar{T}}).\\
\end{split}
\end{equation}
Thus,
\begin{equation}
\begin{split}\label{Wt2}
\lim_{t \rightarrow t^{-}_{s+1}}\!\! W(t)&\leq \lim_{t \rightarrow t^{-}_{s+1}} \!\! \big((1-\frac{\beta_{2}}{\beta_{1}})\textbf{e}^{-\beta_{1}(t -t_{s})}+\frac{\beta_{2}}{\beta_{1}}\big)W(t_{s})\\
&=\big((1-\frac{\beta_{2}}{\beta_{1}})\textbf{e}^{-\beta_{1}(t_{s+1} -t_{s})}+\frac{\beta_{2}}{\beta_{1}}\big)W(t_{s}).\\
%=&\textbf{e}^{-\beta_{1}\bar{T}}+\frac{\beta_{2}}{\beta_{1}}(1-\textbf{e}^{-\beta_{1}\bar{T}}).\\
\end{split}
\end{equation}
Let
 \begin{equation*}\label{rho1}
\begin{split}
&\rho_{s}=(1-\frac{\beta_{2}}{\beta_{1}})\textbf{e}^{-\beta_{1}(t_{s+1}-t_{s})}+\frac{\beta_{2}}{\beta_{1}},~s\in\mathbb{N}\\ &\rho=(1-\frac{\beta_{2}}{\beta_{1}})\textbf{e}^{-\beta_{1}h}+\frac{\beta_{2}}{\beta_{1}}.\\
\end{split}
\end{equation*}
Since $t_{s+1}-t_{s}\geq h$ for all $s\in\mathbb{N}$ and $0<\beta_{2}<\beta_{1}$, we obtain
 \begin{equation}\label{rho2}
\begin{split}
\rho_{s}&\leq\rho=\textbf{e}^{-\beta_{1}h}+\frac{\beta_{2}}{\beta_{1}}(1-\textbf{e}^{-\beta_{1}h})\\
&<\textbf{e}^{-\beta_{1}h}+1-\textbf{e}^{-\beta_{1}h}=1.\\
\end{split}
\end{equation}
Since $W(t)$ is continuous, using  \eqref{Wt2} and \eqref{rho2} gives
 \begin{equation}\label{Wt3}
\begin{split}
W(t_{s+1})=\lim_{t \rightarrow t^{-}_{s+1}} W(t)\leq\rho_{s}W(t_{s})\leq\rho W(t_{s}).
\end{split}
\end{equation}
Therefore, $W(t_{s})$ converges to zero as $s$ tends to infinity, which implies $\lim_{t\rightarrow\infty} W(t)=0$ since $W(t)$ is continuous over $[0,\infty)$.
\end{Proof}
%\begin{Remark}
%It is noted that the proof of Lemma \ref{Lemma1} is inspired by \cite{QianC2012}.
%\end{Remark}

\section{Static Network Case}\label{Static}
In this section, we will first consider the leader-following consensus problem for the multi-agent system composed of \eqref{Follower1} and \eqref{Leader1} under static networks by a distributed sampled-data state feedback control law.

To solve our problem, we need one more assumption on the communication graph as follows:

\begin{Assumption}\label{Ass2}
 Every node $i=1,\cdots,N$ is reachable from node $0$ in the digraph $\mathcal{\bar{G}}$.
\end{Assumption}

\begin{Remark}\label{RemarkAss2}
Assumption \ref{Ass2}  allows the communication graph to be directed, and contains the undirected  graph as a special case. Under this assumption, by Lemma 4 of \cite{HuJ2007}, $H$ is an $\mathcal{M}$-matrix. Then, by Theorem 2.5.3 of \cite{Horn1991}, there exists a positive definite diagonal matrix $D=\mbox{diag}(d_{1},\cdots,d_{N})$ such that $DH+H^{T}D$ is positive definite.
\end{Remark}

%Next, we reorder the sampling time instants of all agents in an ascending sequence as follows:
%\begin{equation}\label{tki1}
%\begin{split}
% \{t_{k}^{i}: i=1,\cdots,N,~k\in\mathbb{N}\}=\{t_{s}: s\in\mathbb{N}\}
%\end{split}
%\end{equation}
%where $0=t_{0}<t_{1}<t_{2}<\cdots$. Clearly, this new time sequence has the following two properties:
%\begin{itemize}
%  \item $\{t_{s}: s\in\mathbb{N}\}$ can be aperiodic, and, for any $s\in\mathbb{N}$,  $t_{s+1}-t_{s}$ is equal to a positive integer multiple of $h$ and $t_{s+1}-t_{s}\leq\min\{T_{1},\cdots,T_{N}\}$.
%  \item For any time interval $[t_{s},t_{s+1})$  with $s\in\mathbb{N}$ and any $i=1,\cdots,N$, there exists a time interval $[t_{k}^{i},t_{k+1}^{i})$ for some $k\in\mathbb{N}$ such that $[t_{s},t_{s+1})\subseteq[t_{k}^{i},t_{k+1}^{i})$.
%\end{itemize}
%According to the second property, the control law \eqref{u1} can be rewritten as follows:
For the static network case, the control law \eqref{u1} can be simplified as follows:
\begin{equation}\label{u3}
\begin{split}
u_{i}(t)\!=\!K\!\sum_{j=0}^{N}\bar{a}_{ij}(x_{j}(t_{s})\!-\!x_{i}(t_{s})),~\forall t\in[t_{s},t_{s+1})
\end{split}
\end{equation}
%where $i=1,\cdots,N$, $t_{s}=sT$, and $s\in\mathbb{N}$.
where $i=1,\cdots,N$, $t_{0}=0$, $t_{s+1}=t_{s}+T_{s}$, $s\in\mathbb{N}$, and $T_{s}\in[\underline{T},\bar{T}]$.
%with $\underline{h}\leq\bar{h}$ being two positive real numbers.

For $i=0,1,\cdots,N$, let
\begin{equation}\label{barxi1}
\begin{split}
&\bar{x}_{i}(t)=x_{i}(t)-x_{0}(t)\\
%\end{split}
%\end{equation}
%\begin{equation}\label{tildex1}
%\begin{split}
&\tilde{x}_{i}(t)=\bar{x}_{i}(t_{s})-\bar{x}_{i}(t),~\forall t\in[t_{s},t_{s+1}).
\end{split}
\end{equation}
Then, according to \eqref{Follower1}, \eqref{Leader1}, \eqref{u3} and \eqref{barxi1}, for $i=1,\cdots,N$, we have
\begin{equation}\label{dotbarxi1}
\begin{split}
\dot{\bar{x}}_{i}(t)=&\dot{x}_{i}(t)-\dot{x}_{0}(t)\\
=&Ax_{i}(t)+BK\sum_{j=0}^{N}\bar{a}_{ij}(x_{j}(t_{s})\!-\!x_{i}(t_{s}))\\
&-Ax_{0}(t)\\
=&A\bar{x}_{i}(t)+BK\sum_{j=0}^{N}\bar{a}_{ij}(\bar{x}_{j}(t_{s})\!-\!\bar{x}_{i}(t_{s}))\\
=&A\bar{x}_{i}(t)+BK\sum_{j=0}^{N}\bar{a}_{ij}(\bar{x}_{j}(t)\!-\!\bar{x}_{i}(t))\\
&+BK\sum_{j=0}^{N}\bar{a}_{ij}(\tilde{x}_{j}(t)\!-\!\tilde{x}_{i}(t)),~\forall t\in[t_{s},t_{s+1}).\\
\end{split}
\end{equation}
Let $\bar{x}=\mbox{col}(\bar{x}_{1},\cdots,\bar{x}_{N})$ and $\tilde{x}=\mbox{col}(\tilde{x}_{1},\cdots,\tilde{x}_{N})$. Then  we further put \eqref{dotbarxi1} into the following compact form:
\begin{equation}\label{dotbarx1}
\begin{split}
\dot{\bar{x}}(t)=&(I_{N}\otimes A)\bar{x}(t)-(H\otimes BK)\bar{x}(t_{s})\\
=&(I_{N}\otimes A-(H\otimes BK))\bar{x}(t)\\
&-(H\otimes BK)\tilde{x}(t),~\forall t\in[t_{s},t_{s+1}).\\
\end{split}
\end{equation}

Since $(A,B)$ is stabilizable and $(I_{n},A)$ is observable, from \cite{Kucera1972}, there exists a unique positive definite matrix $P$ satisfying the following Riccati equation
\begin{equation}\label{Riccati1}
\begin{split}
PA+A^{T}P-\mu_{1}PBB^{T}P+\mu_{2}I_{n}=0
\end{split}
\end{equation}
where $\mu_{1}$ and $\mu_{2}$ are any positive real numbers.

Before giving our main result, we introduce some notation.
 Let
\begin{equation}\label{notation1}
\begin{split}
&d_{m}=\lambda_{\min}(D),~\lambda_{m}=\lambda_{\min}(D\otimes P)\\
&d_{M}=\lambda_{\max}(D),~\lambda_{M}=\lambda_{\max}(D\otimes P)\\
&\lambda_{1}=\lambda_{\min}(DH+H^{T}D),~\alpha_{1}=\frac{\mu_{1}d_{M}}{\lambda_{1}}\\
&\alpha_{2}=2\alpha_{1}\|DH\|\|PBB^{T}P\|,~\alpha_{3}=\frac{\alpha_{2}^{2}}{2d_{m}\mu_{2}}\\
&\alpha_{4}=\frac{(\|A\|+\alpha_{1}\|H\otimes BB^{T}P\|)^{2}}{\lambda_{m}}\\
&c_{1}=\frac{d_{m}\mu_{2}}{2\lambda_{M}},~c_{2}=\alpha_{3}\alpha_{4}.\\
%\frac{\|A\|\!+\!\alpha_{1}\|H\otimes BB^{T}P\|}{\lambda_{m}}.\\
%&d_{m}=\frac{c_{1}}{\lambda_{M}},~d_{2}=2\|PE_{c}\|\frac{\|A_{c}\!-\!E_{c}\|\!+\!\|E_{c}\|}{\lambda_{m}}.\\
%&\bar{d}_{1}=\frac{d_{m}}{d_{2}}\\
\end{split}
\end{equation}
Then we give the following result.
\begin{Theorem}\label{Theorem1}
Under Assumptions \ref{Ass1} and \ref{Ass2}, let $0<\underline{T}\leq\bar{T}<\sqrt{\frac{c_{1}}{c_{2}}}$. Then the distributed sampled-data state feedback control law \eqref{u1} with $K=\alpha_{1}B^{T}P$ and  $T_{s}\in[\underline{T},\bar{T}]$ for all $s\in\mathbb{N}$ solves the leader-following consensus problem for the multi-agent system composed of \eqref{Follower1} and \eqref{Leader1}.
% \begin{equation}\label{T1}
%\begin{split}
%% \nonumber to remove numbering (before each equation)
%T_{i}<\bar{T},~i=1,\cdots,N.%\min\{1,\bar{d}_{1},\bar{d}_{2}\},
%\end{split}
%\end{equation}
\end{Theorem}
\begin{Proof}
First, note that, if $\bar{x}(t_{s})=0$, then, according to \eqref{dotbarx1}, $\bar{x}(t)=0$ for all $t\geq t_{s}$. Thus the problem is obviously solved.

Second, consider the case $\bar{x}(t_{s})\neq0$.
%Let $z=(H\otimes I_{n})\bar{x}$ and $\tilde{z}=(H\otimes I_{n})\tilde{x}$. Then we have
%\begin{equation}\label{dotz1}
%\begin{split}
%\dot{z}(t)=&(H\otimes I_{n})\dot{\bar{x}}(t)\\
%%=&(H\otimes I_{n})\big((I_{N}\otimes A)\bar{x}(t)+(H\otimes BK)\bar{x}(t_{s})\big)\\
%=&(H\otimes I_{n})(I_{N}\otimes A+(H\otimes BK))\bar{x}(t)\\
%&+(H\otimes I_{n})(H\otimes BK)\tilde{x}(t)\\
%=&(H\otimes A+H^{2}\otimes BK))(H^{-1}\otimes I_{n})z(t)\\
%&+(H^{2}\otimes BK)(H^{-1}\otimes I_{n})\tilde{z}(t)\\
%=&(I_{N}\otimes A+H\otimes BK))z(t)\\
%&+(H\otimes BK)\tilde{z}(t),~\forall t\in[t_{s},t_{s+1}).\\
%\end{split}
%\end{equation}
 Let
 \begin{equation}\label{Vbarx1}
\begin{split}
V(\bar{x})=\bar{x}^{T}(D\otimes P)\bar{x}.
\end{split}
\end{equation}
Then we have
 \begin{equation}\label{Vbarx2}
\begin{split}
\lambda_{m}\|\bar{x}\|^{2}\leq V(\bar{x})\leq\lambda_{M}\|\bar{x}\|^{2}.
\end{split}
\end{equation}
Note that $K=\alpha_{1}B^{T}P$. Then, along the trajectory of the closed-loop system \eqref{dotbarx1}, for any $t\in[t_{s},t_{s+1})$, we have
 \begin{equation}\label{dotVbarx1}
\begin{split}
\dot{V}(\bar{x})=&2\bar{x}^{T}(t)(D\otimes P)\dot{\bar{x}}(t)\\
=&2\bar{x}^{T}(t)(D\otimes P)\big((I_{N}\otimes A-(H\otimes BK))\bar{x}(t)\\
&-(H\otimes BK)\tilde{x}(t)\big)\\
=&\bar{x}^{T}(t)\big(D\otimes (PA+A^{T}P)\\
&-\alpha_{1}(DH+H^{T}D)\otimes PBB^{T}P\big)\bar{x}(t)\\
&-2\alpha_{1}\bar{x}^{T}(t)(DH\otimes PBB^{T}P)\tilde{x}(t)\\
\leq&\bar{x}^{T}(t)\big(D\otimes (PA+A^{T}P)\\
&-\mu_{1}D\otimes PBB^{T}P\big)\bar{x}(t)\\
&+2\alpha_{1}\|\bar{x}(t)\|\|DH\|\|PBB^{T}P\|\|\tilde{x}(t)\|\\
=&-\mu_{2}\bar{x}^{T}(t)(D\otimes I_{n})\bar{x}(t)+\alpha_{2}\|\bar{x}(t)\|\|\tilde{x}(t)\|\\
\leq&-d_{m}\mu_{2}\|\bar{x}(t)\|^{2}+\alpha_{2}\|\bar{x}(t)\|\|\bar{x}(t_{s})-\bar{x}(t)\|\\
\leq&-d_{m}\mu_{2}\|\bar{x}(t)\|^{2}+\frac{d_{m}\mu_{2}}{2}\|\bar{x}(t)\|^{2}\\
&+\frac{\alpha_{2}^{2}}{2d_{m}\mu_{2}}\|\bar{x}(t_{s})-\bar{x}(t)\|^{2}\\
=&-\frac{d_{m}\mu_{2}}{2}\|\bar{x}(t)\|^{2}+\alpha_{3}\|\bar{x}(t_{s})-\bar{x}(t)\|^{2}.\\
\end{split}
\end{equation}
Based on  \eqref{dotbarx1} and  \eqref{Vbarx2}, for any $t\in[t_{s},t_{s+1})$,
 \begin{equation}\label{dotbarx2}
\begin{split}
\|\dot{\bar{x}}(t))\|=&\|(I_{N}\otimes A)\bar{x}(t)-(H\otimes BK)\bar{x}(t_{s})\|\\
\leq&\|A\|\|\bar{x}(t)\|+\alpha_{1}\|H\otimes BB^{T}P\|\|\bar{x}(t_{s})\|\\
\leq&\frac{\|A\|}{\sqrt{\lambda_{m}}}\sqrt{V(\bar{x}(t))}\\
&+\frac{\alpha_{1}\|H\otimes BB^{T}P\|}{\sqrt{\lambda_{m}}}\sqrt{V(\bar{x}(t_{s}))}\\
\leq& \frac{\|A\|+\alpha_{1}\|H\otimes BB^{T}P\|}{\sqrt{\lambda_{m}}}\sqrt{V_{M}(t)}\\
\end{split}
\end{equation}
where $V_{M}(t)=\max_{\tau\in[t_{s},t]}V(\bar{x}(\tau))$ for any $t\in[t_{s},t_{s+1})$.
Note that $t_{s+1}-t_{s}=T_{s}\leq \bar{T}$ for any $s\in\mathbb{N}$.
Then, for any $t\in[t_{s},t_{s+1})$,
 \begin{equation}\label{barxts1}
\begin{split}
&\|\bar{x}(t_{s})\!-\!\bar{x}(t)\|\\
\leq&\int_{t_{s}}^{t}\|\dot{\bar{x}}(\tau)\|d\tau\\
\leq&\int_{t_{s}}^{t} \frac{\|A\|+\alpha_{1}\|H\otimes BB^{T}P\|}{\sqrt{\lambda_{m}}}\sqrt{V_{M}(t)}d\tau\\
= &\frac{\|A\|+\alpha_{1}\|H\otimes BB^{T}P\|}{\sqrt{\lambda_{m}}}\sqrt{V_{M}(t)}(t-t_{s})\\
\leq & \frac{\|A\|+\alpha_{1}\|H\otimes BB^{T}P\|}{\sqrt{\lambda_{m}}}\bar{T}\sqrt{V_{M}(t)}\\
\end{split}
\end{equation}
which further implies, for any $t\in[t_{s},t_{s+1})$,
 \begin{equation}\label{barxts2}
\begin{split}
&\|\bar{x}(t_{s})\!-\!\bar{x}(t)\|^{2}\\
\leq & \frac{(\|A\|+\alpha_{1}\|H\otimes BB^{T}P\|)^{2}}{\lambda_{m}}\bar{T}^{2}V_{M}(t)\\
=&\alpha_{4}\bar{T}^{2}V_{M}(t).\\
\end{split}
\end{equation}
According to \eqref{Vbarx2}, \eqref{dotVbarx1} and \eqref{barxts2}, for any $t\in[t_{s},t_{s+1})$, we have
 \begin{equation}\label{dotVbarx2}
\begin{split}
\dot{V}(\bar{x}(t))\!\leq&-\frac{d_{m}\mu_{2}}{2\lambda_{M}}V(\bar{x}(t))+\alpha_{3}\|\bar{x}(t_{s})-\bar{x}(t)\|^{2}\\
\leq&-\frac{d_{m}\mu_{2}}{2\lambda_{M}}V(\bar{x}(t))+\alpha_{3}\alpha_{4}\bar{T}^{2}V_{M}(t)\\
=&-c_{1}V(\bar{x}(t))+c_{2}\bar{T}^{2}V_{M}(t).\\
\end{split}
\end{equation}
Next, we prove
 \begin{equation}\label{Vmax1}
\begin{split}
\max_{\tau\in[t_{s},t]} V(\bar{x}(\tau))=V(\bar{x}(t_{s})), ~ \forall t\in[t_{s},t_{s+1}).
\end{split}
\end{equation}
If \eqref{Vmax1} is not true, then there exists a time instant $t'\in[t_{s},t_{s+1})$ such that $V(\bar{x}(t'))>V(\bar{x}(t_{s}))$. Note that, according to \eqref{dotVbarx1},
\begin{equation}\label{dotVbarxctk1}
\begin{split}
\dot{V}(\bar{x}(t_{s}))\leq-\frac{d_{m}\mu_{2}}{2}\|\bar{x}(t_{s})\|^{2}<0,~\forall \bar{x}(t_{s})\neq0
\end{split}
\end{equation}
which implies that $V(\bar{x}(t))$ will decrease in a short time starting from $t_{s}$. Therefore, there exists another time instant $t''\in[t_{s},t']$ such that
\begin{equation}\label{Vbarxt1}
\begin{split}
&V(\bar{x}(t''))=V(\bar{x}(t_{s}))\\
&\dot{V}(\bar{x}(t''))>0\\
&V(\bar{x}(t))\leq V(\bar{x}(t'')),~\forall t\in[t_{s},t''].\\
\end{split}
\end{equation}
Note that $\bar{T}<\sqrt{\frac{c_{1}}{c_{2}}}$. Then, according to \eqref{dotVbarx2} and the third inequality of \eqref{Vbarxt1}, we have
 \begin{equation}\label{dotVbarx3}
\begin{split}
\dot{V}(\bar{x}(t''))&\leq-c_{1}V(\bar{x}(t''))+c_{2}\bar{T}^{2}V(\bar{x}(t''))<0\\
\end{split}
\end{equation}
which contradicts the second inequality of \eqref{Vbarxt1}. Thus we conclude that \eqref{Vmax1} is true. Then, from \eqref{dotVbarx2}, for any $t\in[t_{s},t_{s+1})$,
 \begin{equation}\label{dotVbarx4}
\begin{split}
\dot{V}(\bar{x}(t))\!\leq&-\!c_{1}V(\bar{x}(t))+c_{2}\bar{T}^{2}V(\bar{x}(t_{s})).\\
\end{split}
\end{equation}

Since $t_{s+1}-t_{s}=T_{s}\geq \underline{T}$ for all $s\in\mathbb{N}$, and $c_{2}\bar{T}^{2}<c_{1}$, by Lemma \ref{Lemma1}, we have $\lim_{t\rightarrow\infty} V(\bar{x}(t))=0$, which implies $\lim_{t\rightarrow\infty} \|\bar{x}(t)\|= 0$.

Thus the proof is complete.
\end{Proof}

%As noted in Remark \ref{RemarkControlLaw1}, the distributed synchronous sampled-data feedback control law \eqref{u2} is a special case of the distributed asynchronous sampled-data feedback control law \eqref{u1}. Then we obtain the following corollary directly.
%
%\begin{Corollary}\label{Corollary1}
%Under Assumptions \ref{Ass1} and \ref{Ass2}, let $\bar{T}<\sqrt{\frac{c_{1}}{c_{2}}}$. Then the distributed synchronous sampled-data state feedback control law \eqref{u2} with $K=\alpha_{1}B^{T}P$ and  $T\leq\bar{T}$  solves the leader-following consensus problem for the multi-agent system composed of \eqref{Follower1} and \eqref{Leader1}.
% %with the static digraph $\mathcal{\bar{G}}$.
%\end{Corollary}

\begin{Remark}\label{RemarkTheorem1a}
In fact, it is possible to design a control law and an upper bound independent of the specific connection information of the graph. Since the number of all graphs with a finite number of nodes is  finite, we can calculate all possible $H$ and hence $D$ off-line.  For this purpose, let $\mathcal{J}=\{1,\cdots,N_{0}\}$, where $N_{0}$ is the total number of all connected graphs with the  number of the nodes equal to $N+1$. Then, all the parameters defined in \eqref{notation1} can also be calculated off-line as follows:
\begin{equation}\label{notation1b}
\begin{split}
&d_{m}=\min_{j\in\mathcal{J}}\{\lambda_{\min}(D_{j})\},~\lambda_{m}=\min_{j\in\mathcal{J}}\{\lambda_{\min}(D_{j}\otimes P)\}\\
&d_{M}=\max_{j\in\mathcal{J}}\{\lambda_{\max}(D_{j})\},~\lambda_{M}\!=\!\max_{j\in\mathcal{J}}\{\lambda_{\max}(D_{j}\otimes P)\}\\
&\lambda_{1}=\min_{j\in\mathcal{J}}\{\lambda_{\min}(D_{j}H_{j}+H_{j}^{T}D_{j})\},~\alpha_{1}=\frac{\mu_{1}d_{M}}{\lambda_{1}}\\
&\alpha_{2}=\max_{j\in\mathcal{J}}\{2\alpha_{1}\|D_{j}H_{j}\|\|PBB^{T}P\|\},~\alpha_{3}=\frac{\alpha_{2}^{2}}{2d_{m}\mu_{2}}\\
&\alpha_{4}=\max_{j\in\mathcal{J}}\{\frac{(\|A\|+\alpha_{1}\|H_{j}\otimes BB^{T}P\|)^{2}}{\lambda_{m}}\}\\
&c_{1}=\frac{d_{m}\mu_{2}}{2\lambda_{M}},~c_{2}=\alpha_{3}\alpha_{4}.\\
%\frac{\|A\|\!+\!\alpha_{1}\|H\otimes BB^{T}P\|}{\lambda_{m}}.\\
%&d_{m}=\frac{c_{1}}{\lambda_{M}},~d_{2}=2\|PE_{c}\|\frac{\|A_{c}\!-\!E_{c}\|\!+\!\|E_{c}\|}{\lambda_{m}}.\\
%&\bar{d}_{1}=\frac{d_{m}}{d_{2}}\\
\end{split}
\end{equation}
With the parameters given by \eqref{notation1b}, the control law \eqref{u1} applies to all connected graphs with the number of the nodes equal to $N+1$.
Nevertheless, it should be noted that the parameters defined in \eqref{notation1b} are more conservative than those defined in  \eqref{notation1}.
\end{Remark}

\section{Switching Network Case}\label{Switching}
In this section, we will further consider the leader-following consensus problem for the multi-agent system composed of \eqref{Follower1} and \eqref{Leader1} under switching networks by a distributed sampled-data state feedback control law.

For this purpose, we introduce another assumption on the communication graph  as follows:

\begin{Assumption}\label{Ass3}
 For any $p\in\mathcal{P}$,  every node $i=1,\cdots,N$ is reachable from node $0$ in the digraph $\mathcal{\bar{G}}_{p}$ and there exists a positive definite diagonal matrix $D=\mbox{diag}(d_{1},\cdots,d_{N})$ such that $DH_{p}+H_{p}^{T}D$ is positive definite.
\end{Assumption}

\begin{Remark}\label{RemarkAss3}
Clearly,  Assumption \ref{Ass3} contains Assumption \ref{Ass2} as a special case.  Next, define a subgraph $\mathcal{G}_{p}=(\mathcal{V},\mathcal{E}_{p})$, where $\mathcal{V}=\{1,\cdots,N\}$ and $\mathcal{E}_{p}\subseteq\mathcal{V}\times\mathcal{V}$ is obtained from $\mathcal{\bar{E}}_{p}$ by removing all edges between node $0$ and the nodes in $\mathcal{V}$. Then Assumption \ref{Ass3} also contains the following assumption as a special case:
\emph{ For any $p\in\mathcal{P}$,  every node $i=1,\cdots,N$ is reachable from node $0$ in the digraph $\mathcal{\bar{G}}_{p}$ and the subgraph $\mathcal{G}_{p}$ is undirected.} This assumption has been used in \cite{HongY2008}. In fact, in some cases, even if the subgraph $\mathcal{G}_{p}$ is directed, it is still possible to find a common diagonal matrix $D$ such that $DH_{p}+H_{p}^{T}D$ is positive definite. One example is given in Section VI-Case B, where the two subgraphs $\mathcal{G}_{1}$ and $\mathcal{G}_{2}$ are directed and a common $D$ still exists.
\end{Remark}

%Next, we also reorder the sampling time instants of all agents in an ascending sequence as \eqref{tki1}.
%Then, under the switching digraph $\mathcal{\bar{G}}_{\sigma(t)}$,  the control law \eqref{u1} can be rewritten as follows:
%\begin{equation}\label{u4}
%\begin{split}
%u_{i}(t)\!=\!K\!\sum_{j=0}^{N}\bar{a}_{ij}(t_{s})(x_{j}(t_{s})\!-\!x_{i}(t_{s})),~\forall t\in[t_{s},t_{s+1})
%\end{split}
%\end{equation}
%where $\bar{a}_{ij}(t_{s})=\bar{a}_{ij}(t_{k}^{i})$ and $x_{j}(t_{s})=x_{j}(t_{k}^{i})$ for $j=0,1,\cdots,N$ and some $k\in\mathbb{N}$.

Let $\bar{x}(t)$ and $\tilde{x}(t)$ be defined as in Section \ref{Static}. Then, for the switching network case, we have
\begin{equation}\label{dotbarx3}
\begin{split}
\dot{\bar{x}}(t)=&(I_{N}\otimes A-(H_{\sigma(t_{s})}\otimes BK))\bar{x}(t)\\
&-(H_{\sigma(t_{s})}\otimes BK)\tilde{x}(t),~\forall t\in[t_{s},t_{s+1}).\\
\end{split}
\end{equation}
Let
\begin{equation}\label{notation2}
\begin{split}
%&\lambda_{m}=\lambda_{\min}(D\otimes P),~\lambda_{M}=\lambda_{\max}(D\otimes P)\\
&\lambda_{1}=\min_{p\in\mathcal{P}}\{\lambda_{\min}(DH_{p}+H_{p}^{T}D)\}\\%,~\alpha_{1}=\frac{\mu_{1}d_{M}}{\lambda_{1}}\\
&\alpha_{2}=\max_{p\in\mathcal{P}}\{2\alpha_{1}\|DH_{p}\|\|PBB^{T}P\|\}\\%,~c_{1}=\frac{d_{m}\mu_{2}}{\lambda_{M}}\\
&\alpha_{4}=\max_{p\in\mathcal{P}}\{\frac{(\|A\|+\alpha_{1}\|H_{p}\otimes BB^{T}P\|)^{2}}{\lambda_{m}}\}\\
%&c_{2}=\max_{p\in\mathcal{P}}\{\alpha_{2}\frac{\|A\|\!+\!\alpha_{1}\|H_{p}\otimes BB^{T}P\|}{\lambda_{m}}\}.\\
%&d_{m}=\frac{c_{1}}{\lambda_{M}},~d_{2}=2\|PE_{c}\|\frac{\|A_{c}\!-\!E_{c}\|\!+\!\|E_{c}\|}{\lambda_{m}}.\\
%&\bar{d}_{1}=\frac{d_{m}}{d_{2}}\\
\end{split}
\end{equation}
The matrix $P$ is  defined as in \eqref{Riccati1} and the other parameters $d_{m},d_{M},\lambda_{m},\lambda_{M},\alpha_{1},\alpha_{3},c_{1},c_{2}$ are defined as in  \eqref{notation1}.
Then we give the following result.

\begin{Theorem}\label{Theorem2}
Under Assumptions \ref{Ass1} and \ref{Ass3}, let $0<\underline{T}\leq\bar{T}<\sqrt{\frac{c_{1}}{c_{2}}}$. Then the distributed sampled-data state feedback control law \eqref{u1} with $K=\alpha_{1}B^{T}P$ and  $T_{s}\in[\underline{T},\bar{T}]$ for all $s\in\mathbb{N}$ solves the leader-following consensus problem for the multi-agent system composed of \eqref{Follower1} and \eqref{Leader1}.
% \begin{equation}\label{T1}
%\begin{split}
%% \nonumber to remove numbering (before each equation)
%T_{i}<\bar{T},~i=1,\cdots,N.%\min\{1,\bar{d}_{1},\bar{d}_{2}\},
%\end{split}
%\end{equation}
\end{Theorem}
\begin{Proof}
The proof is similar to the proof of Theorem \ref{Theorem1}. Choose the same function $V(\bar{x})=\bar{x}^{T}(D\otimes P)\bar{x}$ as in \eqref{Vbarx1}. Note that, under the switching digraph $\mathcal{\bar{G}}_{\sigma(t)}$, $V(\bar{x}(t))$ is still continuous. However, the time derivative of $V(\bar{x}(t))$ is discontinuous not only at the sampling time instants but also at the switching time instants. Nevertheless, with $\lambda_{1}$, $\alpha_{2}$ and $\alpha_{4}$ being defined in \eqref{notation2},  the time derivative of $V(\bar{x}(t))$ satisfies
 \begin{equation}\label{dotVbarx5}
\begin{split}
\dot{V}(\bar{x})=&\bar{x}^{T}(t)\big(D\otimes (PA+A^{T}P)\\
&-\alpha_{1}(DH_{\sigma(t_{s})}+H_{\sigma(t_{s})}^{T}D)\otimes PBB^{T}P\big)\bar{x}(t)\\
&-2\alpha_{1}\bar{x}^{T}(t)(DH_{\sigma(t_{s})}\otimes PBB^{T}P)\tilde{x}(t)\\
\leq&\bar{x}^{T}(t)\big(D\otimes (PA+A^{T}P)\\
&-\mu_{1}D\otimes PBB^{T}P\big)\bar{x}(t)\\
&+2\alpha_{1}\|\bar{x}(t)\|\|DH_{\sigma(t_{s})}\|\|PBB^{T}P\|\|\tilde{x}(t)\|\\
\leq&-\mu_{2}\bar{x}^{T}(t)(D\otimes I_{n})\bar{x}(t)+\alpha_{2}\|\bar{x}(t)\|\|\tilde{x}(t)\|\\
\leq&-d_{m}\mu_{2}\|\bar{x}(t)\|^{2}+\alpha_{2}\|\bar{x}(t)\|\|\bar{x}(t_{s})-\bar{x}(t)\|\\
\leq&-\frac{d_{m}\mu_{2}}{2}\|\bar{x}(t)\|^{2}+\alpha_{3}\|\bar{x}(t_{s})-\bar{x}(t)\|^{2}\\
\end{split}
\end{equation}
for all $t\in[t_{s},t_{s+1})$. The remaining part of the proof is the same as that in the proof of Theorem \ref{Theorem1}.
\end{Proof}

%Similarly, we have the following corollary.
%\begin{Corollary}\label{Corollary2}
%Under Assumptions \ref{Ass1} and \ref{Ass3}, let $\bar{T}<\sqrt{\frac{c_{1}}{c_{2}}}$. Then the distributed synchronous sampled-data state feedback control law \eqref{u2} with $K=\alpha_{1}B^{T}P$ and  $T\leq\bar{T}$  solves the leader-following consensus problem for the multi-agent system composed of \eqref{Follower1} and \eqref{Leader1}.
%%with the switching digraph $\mathcal{\bar{G}}_{\sigma(t)}$.
%\end{Corollary}

%\begin{Remark}\label{RemarkTheorem2a}
%Similarly, in practice, if each agent does not know the specific matrix $H_{p}$, then all the parameters  can be calculated off-line as in \eqref{notation1b}.
%\end{Remark}
\begin{Remark}\label{RemarkTheorem2b}
The  upper bound for the sampling intervals given in Theorems \ref{Theorem1} and \ref{Theorem2} may be  conservative. In practice, even if the sampling intervals are greater than the given upper bound,  the problem may still be solved by the proposed control law.
\end{Remark}

\begin{Remark}\label{RemarkTheorem2c}
References \cite{XiaoF2008} and \cite{GaoY2011} also studied the sampled-data consensus problem, where the communication graph condition is weaker than Assumption \ref{Ass3} and  the time delay issue was considered in \cite{XiaoF2008}. Nevertheless, there are at least  four main differences or novelties between the results in this paper and the results in \cite{XiaoF2008} and \cite{GaoY2011}. First, references \cite{XiaoF2008} and \cite{GaoY2011} considered the sampled-data leaderless consensus problem, whereas we consider the sampled-data leader-following consensus problem.
  Second, references \cite{XiaoF2008} and \cite{GaoY2011} considered  single integrator systems and double integrator systems, respectively, whereas we consider a class of general linear multi-agent systems, which contains single integrator systems and double integrator systems as special cases.
Third,  in  \cite{XiaoF2008} and \cite{GaoY2011}, the problem was transformed into the asymptotic stability problem of a discrete-time system, whereas we develop a new technical lemma to analyze the stability of the piecewise-continuous closed-loop system directly. Finally, we give an explicit upper bound for the sampling intervals that guarantees the stability and performance of the closed-loop system as long as all the sampling intervals are smaller than this upper bound.
\end{Remark}

%\begin{Remark}
%It is of interest to compare our results with some existing results in \cite{DingL2013,Rakkiyappan2015,ZhangD2018}, where  the SDLFCP and the sampled-data leader-following mean square consensus problem for general linear multi-agent systems have been studied. Nevertheless, there are three main differences between our results with these existing results. First,  the sampled-data control laws in \cite{DingL2013,Rakkiyappan2015,ZhangD2018} are updated synchronously, while here our control law is updated asynchronously, which is easier to be implemented in practice. Second, the communication network is assumed to be static in \cite{DingL2013,Rakkiyappan2015,ZhangD2018}, while here both the static directed network case and the switching directed network case are considered. Third, the solvability conditions of the problems in \cite{DingL2013,Rakkiyappan2015,ZhangD2018} were converted into the solvability conditions of some complex linear matrix inequalities. However, here we show that the problem can be solved by the proposed control law as long as the sampling periods of all agents are smaller than an explicitly given threshold.
%\end{Remark}

\section{An Example}\label{Example}
In this section, we consider a linear multi-agent system with the leader system as follows:
 \begin{equation}\label{Leader2}
\begin{split}
\dot{x}_{0}=\left[
              \begin{array}{cc}
                -0.38 & 0.72 \\
                -0.68 & 0.42 \\
              \end{array}
            \right]x_{0}
\end{split}
\end{equation}
 and the four follower systems as follows:
 \begin{equation}\label{Follower2}
\begin{split}
\dot{x}_{i}=\left[
              \begin{array}{cc}
                -0.38 & 0.72 \\
                -0.68 & 0.42 \\
              \end{array}
            \right]x_{i}+\left[
                           \begin{array}{c}
                             0.26 \\
                             0.31 \\
                           \end{array}
                         \right]u_{i}
\end{split}
\end{equation}
for $i=1,2,3,4$. Clearly, Assumption \ref{Ass1} is satisfied.
\subsection{Static Network Case}
\begin{figure}[H]
\centering
\includegraphics[scale=0.55]{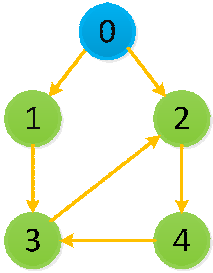}
\caption{Communication graph $\mathcal{\bar{G}}$.} \label{g1}
\end{figure}

Consider the static communication graph $\mathcal{\bar{G}}$ in Figure \ref{g1}, where node $0$ is associated with the leader system, and the other nodes are associated with the follower systems. It is easy to see that Assumption \ref{Ass2} is satisfied and
 \begin{equation*}%\label{Leader2}
\begin{split}
H=\left[
  \begin{array}{cccc}
    1 & 0 & 0 & 0 \\
    0 & 2 & -1 & 0 \\
    -1 & 0 & 2 & -1 \\
    0 & -1 & 0 & 1 \\
  \end{array}
\right].
\end{split}
\end{equation*}
Choose $D=I_{4}$. Then it is easy to check that $DH+H^{T}D$ is positive definite. Choose $\mu_{1}=\mu_{2}=1$. Then solving  \eqref{Riccati1} gives
 \begin{equation*}%\label{Leader2}
\begin{split}
P=\left[
  \begin{array}{cc}
    7.2138 &  -3.6897 \\
   -3.6897 &   6.3388\\
  \end{array}
\right].
\end{split}
\end{equation*}
Following the procedures described in Section \ref{Static}, we obtain $\bar{T}=0.0186$ and $K=\left[
                                                                                                  \begin{array}{cc}
                                                                                                    0.8874 &  1.2195 \\
                                                                                                  \end{array}
                                                                                                \right]
$. We further choose $\underline{T}=0.001$. Then, by Theorem \ref{Theorem1}, the distributed sampled-data state feedback control law \eqref{u1} with $K=\left[
                                                                                                  \begin{array}{cc}
                                                                                                    0.8874 &  1.2195 \\
                                                                                                  \end{array}
                                                                                                \right]$
and $T_{s}\in[0.001,0.0186]$ for all $s\in\mathbb{N}$ solves the leader-following consensus problem for the multi-agent system composed of \eqref{Leader2} and \eqref{Follower2}.

Simulation is performed with $T_{s}=l_{s}h$, $h=0.001$, $l_{s}$ randomly chosen in the set $\mathcal{S}_{1}=\{1,2,\cdots,18\}$, $s\in\mathbb{N}$, and
 \begin{equation*}%\label{Leader2}
\begin{split}
&x_{0}=[1.2, -0.8]^{T}\\
&x_{1}=[2.4, -1.6]^{T},~x_{2}=[-1.4, 2.6]^{T}\\
&x_{3}=[1.8,-2.5]^{T},~x_{4}=[-0.2, 1.3]^{T}.\\
\end{split}
\end{equation*}

\begin{figure}[H]
\centering
\includegraphics[scale=0.55]{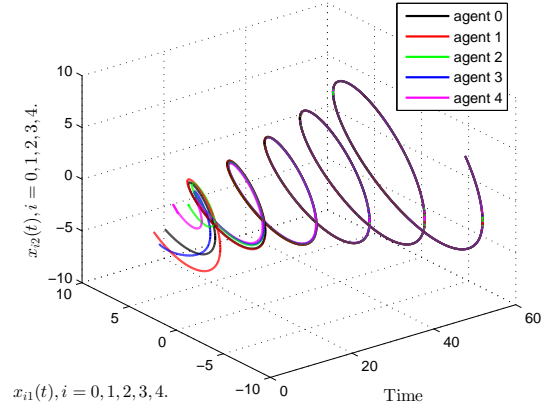}
\caption{Trajectories of all agents under static network.} \label{Trajectory1}
\end{figure}

\begin{figure}[H]
\centering
\includegraphics[scale=0.55]{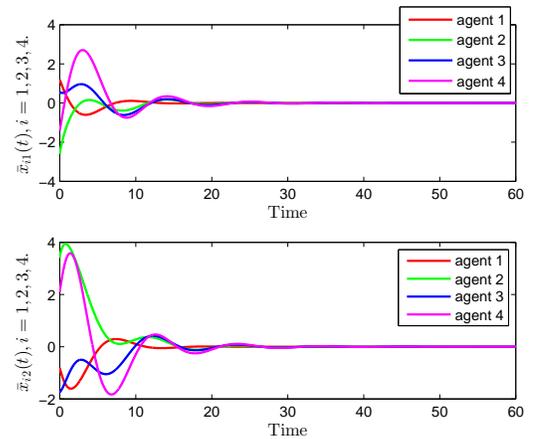}
\caption{Tracking errors of all agents under static network.} \label{Error1}
\end{figure}

The trajectories and tracking errors of all agents under the communication graph $\mathcal{\bar{G}}$ are shown in Figure \ref{Trajectory1} and Figure \ref{Error1}, respectively. It can be found that the trajectories of all follower systems approach the trajectory of the leader system asymptotically, and thus the tracking errors of all agents approach zero asymptotically. Therefore,  the leader-following consensus is achieved satisfactorily.


\subsection{Switching Network Case}
\begin{figure}[H]
  \centering
  \subfigure[$\bar{\mathcal{G}}_{1}$]{
    %\label{g1} %% label for first subfigure
    \includegraphics[scale=0.55]{g1.eps}}
  \hspace{0.5in}
  \subfigure[$\bar{\mathcal{G}}_{2}$]{
    %\label{g2} %% label for second subfigure
    \includegraphics[scale=0.55]{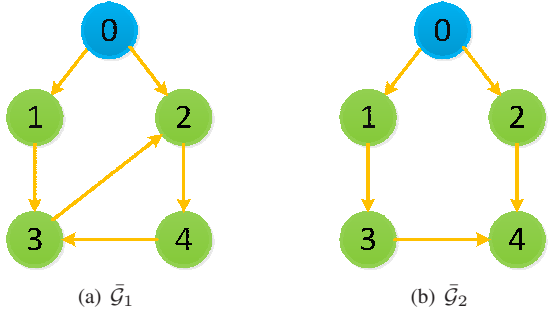}}
  \caption{Two communication graphs $\bar{\mathcal{G}}_{1}$ and $\bar{\mathcal{G}}_{2}$.}
  \label{g} %% label for entire figure
\end{figure}

Consider the switching communication graph $\mathcal{\bar{G}}_{\sigma(t)}$, where
\begin{equation}\label{sigmat}
\begin{split}
\sigma(t)= \left\{
  \begin{array}{ll}
    1, & \hbox{if\ $lT_{0}\leq t< (l+\dfrac{2}{3})T_{0}$} \\
    2, & \hbox{if\ $(l+\dfrac{2}{3})T_{0}\leq t< (l+1)T_{0}$} \\
  \end{array}
\right.
 \end{split}
\end{equation}
for $l=0,1,2,\cdots$ and $T_{0}=1$, and the two communication graphs are described  in Figure \ref{g}. It is easy to obtain
 \begin{equation*}%\label{Leader2}
\begin{split}
H_{1}\!=\!\!\left[\!\!
  \begin{array}{cccc}
    1 & 0 & 0 & 0 \\
    0 & 2 & -1 & 0 \\
    -1 & 0 & 2 & -1 \\
    0 & -1 & 0 & 1 \\
  \end{array}
\!\!\right],~H_{2}\!=\!\!\left[\!\!
  \begin{array}{cccc}
    1 & 0 & 0 & 0 \\
    0 & 1 & 0 & 0 \\
    -1 & 0 & 1 & 0 \\
    0 & -1 & -1 & 2 \\
  \end{array}
\!\!\right].
\end{split}
\end{equation*}
Choose $D=I_{4}$. Then it is ready to check that $DH_{1}+H_{1}^{T}D$ and $DH_{2}+H_{2}^{T}D$ are both positive definite. Thus Assumption \ref{Ass3} is also satisfied.

 Choosing $\mu_{1}=\mu_{2}=1$, and following the procedures described in Section \ref{Switching}, we obtain $\bar{T}=0.0167$ and $K=\left[
                                                                                                  \begin{array}{cc}
                                                                                                    0.9483 &  1.3033 \\
                                                                                                  \end{array}
                                                                                                \right]
$. We further choose $\underline{T}=0.001$.
Then, by Theorem \ref{Theorem2}, the distributed sampled-data state feedback control law \eqref{u1} with $K=\left[
                                                                                                  \begin{array}{cc}
                                                                                                    0.9483 &  1.3033 \\
                                                                                                  \end{array}
                                                                                                \right]$
and $T_{s}\in[0.001,0.0167]$ for all $s\in\mathbb{N}$ solves the leader-following consensus problem for the multi-agent system composed of \eqref{Leader2} and \eqref{Follower2}.

\begin{figure}[H]
\centering
\includegraphics[scale=0.55]{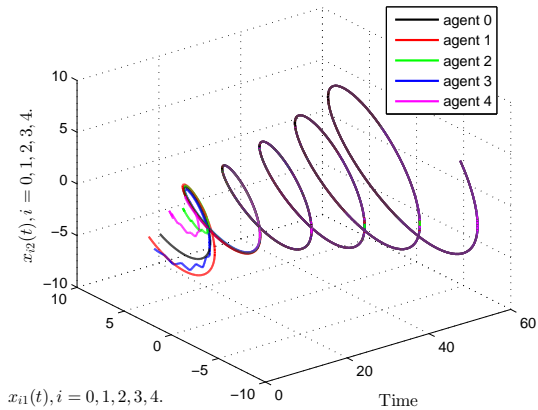}
\caption{Trajectories of all agents under switching network.} \label{Trajectory2}
\end{figure}
\begin{figure}[H]
\centering
\includegraphics[scale=0.55]{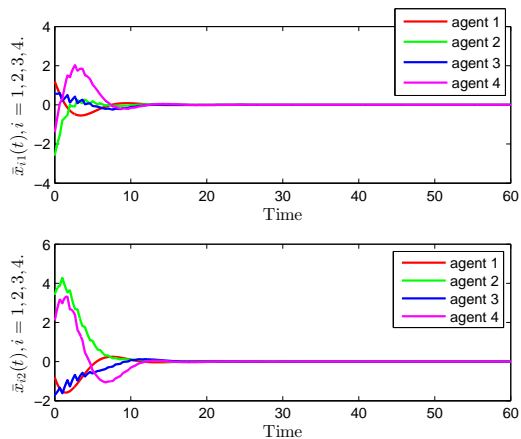}
\caption{Tracking errors of all agents under switching network.} \label{Error2}
\end{figure}

Simulation is performed with $T_{s}=l_{s}h$, $h=0.001$, $l_{s}$ randomly chosen in the set $\mathcal{S}_{2}=\{1,2,\cdots,16\}$, $s\in\mathbb{N}$, and the same initial states as those for the static network case.
The trajectories and tracking errors of all agents under the switching communication graph $\mathcal{\bar{G}}_{\sigma(t)}$ are shown in Figure \ref{Trajectory2} and Figure \ref{Error2}, respectively. As expected,  the trajectories of all follower systems approach the trajectory of the leader system asymptotically, and thus the tracking errors of all agents approach zero asymptotically. Therefore,  the leader-following consensus is achieved satisfactorily.

\section{Conclusion}\label{Conclusion}
In this paper, we have studied  the sampled-data leader-following consensus problem for a class of general linear multi-agent systems. Both the static network case and the switching network case have been studied. It has been shown that the problem can be solved by the proposed distributed sampled-data control law if all the sampling intervals are smaller than an explicitly given threshold.

It would be interesting to further consider the sampled-data leader-following consensus problem for linear multi-agent systems with time delay, parameter uncertainties, and to weaken the condition on communication topologies. The results of this paper and some existing results in \cite{HanD2013,XiaoF2008,ZhangL2013} may shed some light on this future work.

\end{document}